\documentclass{optica-article}

\journal{oe}

\articletype{Research Article}
\usepackage{amsfonts, amsmath, bm, bbm, graphicx, epsfig, epstopdf}
\usepackage{dcolumn}
\usepackage{textcomp}
 \usepackage[caption=false]{subfig}
 \usepackage[caption=false]{subfig}
\usepackage{dsfont}

 \usepackage[caption=false]{subfig}
 \usepackage[caption=false]{subfig}
 
\usepackage{mathtools}
\usepackage{lipsum}

    \usepackage{braket}
\renewcommand\bra[1]{{\langle{#1}|}}
\makeatletter
\renewcommand\ket[1]{%
  \@ifnextchar\bra{\k@t{#1}\!}{\k@t{#1}}%
}
\newcommand\k@t[1]{{|{#1}\rangle}}
\makeatother

\newcommand{\appropto}{\mathrel{\vcenter{
  \offinterlineskip\halign{\hfil$##$\cr
    \propto\cr\noalign{\kern2pt}\sim\cr\noalign{\kern-2pt}}}}}

\begin{document}

\title{ All-optical quantum information processing via a single-step Rydberg blockade gate}
%\title{ All-optical quantum information processing via a single-step  Rydberg geometric  gate}

\author{Mohammadsadegh Khazali}
\affil{ Institute for Research in Fundamental Sciences (IPM), Tehran 19538-33511, Iran\\
 Department of Physics, University of Tehran, 14395-547, Tehran, Iran\\
Institute for Quantum Optics and Quantum Information of the Austrian Academy of Sciences, A-6020 Innsbruck, Austria}

\begin{abstract}
One of the critical elements in the realization of the quantum internet are deterministic two-photon gates. This $CZ$ photonic gate also completes a set of universal gates for all-optical quantum information processing. This article discusses an approach to realize high fidelity $CZ$ photonic gate by storing both control and target photons within an atomic ensemble using non-Rydberg  electromagnetically induced transparency (EIT) followed by a fast, single-step  Rydberg excitation with global lasers. The proposed  scheme operates by relative intensity modulation of two lasers used in Rydberg excitation. Circumventing the conventional $\pi$-gap-$\pi$ schemes,  the proposed operation features continuous laser protection of the Rydberg atoms from the environment noise.  The complete spatial overlap of stored photons inside the blockade radius  optimizes the optical depth and simplifies the experiment. The coherent operation here is performed in the region that was  dissipative in the previous Rydberg EIT schemes. Encountering the main imperfection sources, i.e. the spontaneous emission of the Rydberg and intermediate levels, population rotation errors, Doppler broadening of the transition lines, storage/retrieval efficiency, and atomic thermal motion induced decoherence, this article concludes that with realistic experimental parameters  99.7\% fidelity is achievable.  
\end{abstract}

%\setboolean{displaycopyright}{true}

%\begin{document}

%\maketitle
\section{Introduction}

While the non-interacting features of photons make them an ideal platform for quantum information, the same characteristic deprives two-body deterministic gates. These gates are required for efficient long-distance quantum communication. They are also needed for simple operations in the nodes of quantum internet \cite{Kim08}. 
In contrast to photons, highly excited Rydberg atoms feature controllable long-range interaction ideal for the implementation of quantum information \cite{Saf10,Lev19,Kha20,Kha17,Khaz21,Khaz22,Khaz21IJAP} and quantum matters \cite{Kha16,Kha21,Kha22,Kha23}.
Hence, one approach in making two-qubit photonic gates is to map the quantum state of non-interacting photons to the highly interacting Rydberg atoms \cite{Bri03,Par14,Hac16,Hao15,Kha15,Das16,Wad16,Lah17,Tho17,Kha19,Tia19}.  
 This could be realized under the concept of Rydberg electromagnetically induced transparency (EIT), which supports the lossless propagation of single photons in the form of the so-called Rydberg polaritons \cite{Kha19,Fri05,Gor11,Pri10,Fir13,Bau14,Gor14,Tia14,Tia16,Nin16,Bus17,Mur17,Tom17}. However, by having two Rydberg polaritons, the influence of Rydberg blockade breaks the underlying EIT, leading to the loss of photons \cite{He14,Kha19}.

\begin{figure*}[h]
\centering 
\scalebox{0.4}{\includegraphics{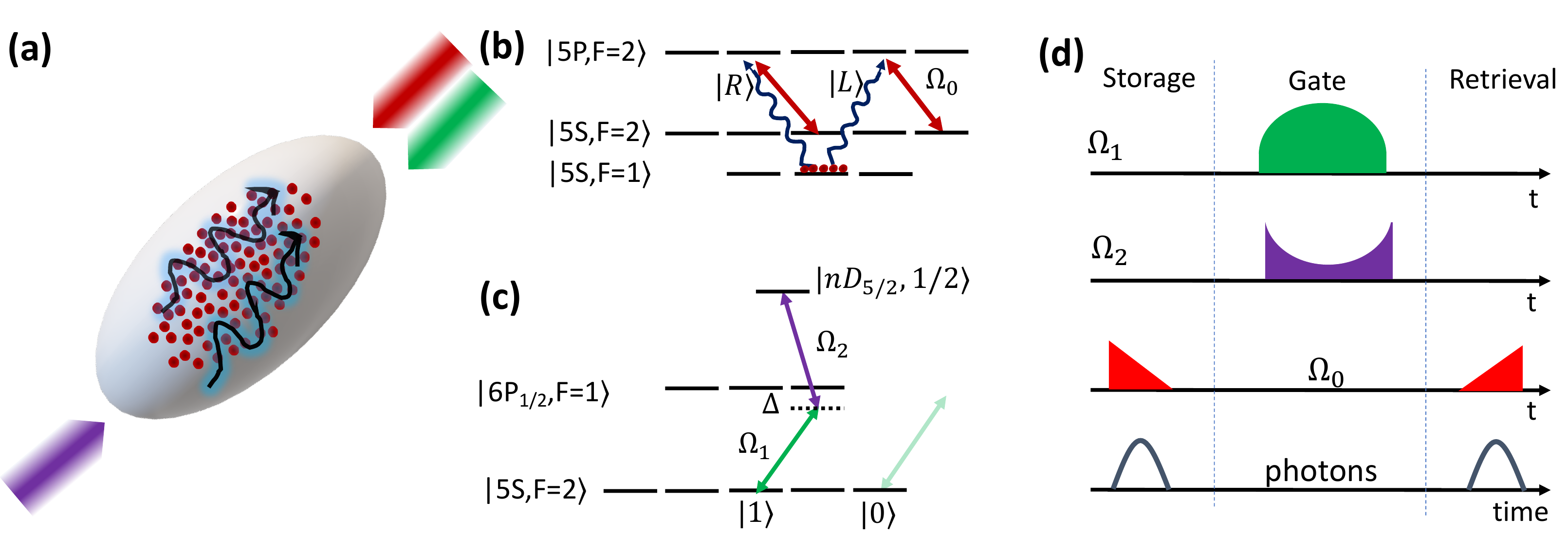}} 
\caption{level scheme. (a) Both control and target photons are stored in the same atomic ensemble that is smaller than the blockade volume. (b) The photon storage is applied via non-Rydberg EIT. The qubit states $|R\rangle$ and $|L\rangle$ would be stored in different ground hyperfine states. (c) Rydberg excitation is derived in a  two-photon scheme with global lasers, where the relative strength of two lasers $\Omega_1/\Omega_2$ varies over time. Selective excitation of the $|\bar{1}\rangle$ qubit states are governed by using $\sigma^{+}$ circular polarization for $\Omega_1$ laser.  (d) The operation timing shows the sequence of applying laser pulses for the photon storage/retrieval and Rydberg gate operation with optimized laser intensity time-varying profiles.}
\label{Fig1}
\end{figure*}

To circumvent this decoherence source, new ideas are being explored that go beyond this blockade-type nonlinearity in Rydberg-EIT systems and allow for coherent effective polariton interactions. One approach was to change the  interaction strength over the storage and gate operation periods \cite{Par14}. This approach required the Rydberg polaritons to be stored  out of the  small blockade radius and at the same time be inside  the large blockade radius for the gate operation. The competing conditions would affect the high-fidelity operation. Also storing the polaritons in the appropriate place turned out to be challenging. The other approach was to use exchange interaction as the source of conditional phase instead of blockade \cite{Tho17}. In this case, at large optical depths (OD), the operation occurs before the polaritons enter the blockade radius. In a single-lane collision, due to the blockade effect, the optical depth of the medium would depend on the two-photon qubit configurations. The conditional optical depth in this case reduces the fidelity.  
 Hence, two rail collision is studied \cite{Kha19} to suppress the loss and conditional OD. However, the competition between the transparency and exchange fidelity still acts as the limiting factor. Even outside the blockade radius, it is argued that the pairwise interaction among Rydberg polaritons could still decouple them from the EIT control field \cite{Tre15}. 
The other approach studies the storage of photons out of the blockade radius via normal EIT followed by Rydberg excitation \cite{Kha15}. The conditional phase in that scheme is accumulated over time by the space-dependent Rydberg interaction. The accumulated in-homogenous phase, suppresses the fidelity by entangling the spatial and computational basis and reordering the phase matching direction. The common decoherence channel in the aforementioned  Rydberg gate schemes was the $\pi$-gap-$\pi$ operation timeline where the excited atoms remained in the Rydberg level over a period of time, unprotected from the laser. This would prevent successful retrieval of the qubit basis \cite{Mal15,Gra19}.

This paper studies a case where storage and interaction are performed in different stages.  Both control and target photons are stored in the same medium on top of each other via non-Rydberg EIT. The non-Rydberg EIT \cite{Fle00,Kav13,Wan19} is the most successful approach in storage and retrieval of single photons in atomic excitations.
The absence of blockade among non-Rydberg polaritons provides unconditional uniform storage and retrieval efficiency for all qubit configurations. The perfect overlap of both spin waves optimizes the blockade mechanism and improves the storage region for photons, enhancing the phase-matching direction. Over the gate operation, the new scheme operates by a fast, continuous, and global laser pulse, exciting both spin-waves in $\ket{\bar{1}}$ logical state to the Rydberg level. The relative phase arrangement and retrieval of the logical basis in different qubit configurations are obtained by modulating the relative intensity of two Rydberg exciting lasers combined with the conditional dipolar interaction.  This arrangement would close a major decoherence source in conventional $\pi$-gap-$\pi$ Rydberg gate schemes where the operation steps contain a period where excited atoms remain in the sensitive Rydberg level, exposed to the stray fields without laser protection \cite{Mal15,Gra19}. Considering different decoherence sources, we find the optimum ensemble size and laser arrangement that lead to homogenous conditional phase  over the spin-waves, and large OD for efficient storage and retrieval of photons. The total fidelity of 99.7\% is achievable with experimentally realistic parameters.

\section{ Gate Scheme}

 The photonic qubits are encoded in the left $\ket{L}$ and right $\ket{R}$ circular polarization components. The {\it photon storage process} is performed via non-Rydberg EIT \cite{Fle00} in the same medium with the level scheme that is shown in Fig.~\ref{Fig1}a. The atomic ensemble is initialized in the long-lived hyperfine ground state $\ket{g}=\ket{5S_{1/2},F=1,m_f=0}$.  Different polarization components of a photon would be stored in distinguished ground hyperfine states $\ket{0}=\ket{F=2,m_f=2}$ and $\ket{1}=\ket{2,0}$. A single photon forms a single excitation that is delocalized over the atomic ensemble.  The spin-wave qubit could be written as  
\begin{equation}
C_0\ket{\bar{0}}+C_1\ket{\bar{1}}=1/\sqrt{N}\sum_i (C_0({\bf x}_i)e^{i{\bf k}.{\bf x}_i}\sigma_{0g}^{(i)}+C_1({\bf x}_i)e^{i{\bf k}.{\bf x}_i}\sigma_{1g}^{(i)})\ket{G}
\end{equation}
 where $\sigma^{(i)}_{lm}=\ket{l}\bra{m}$ is the transition operator of the $i^{th}$ atom, the overline indicates the delocalized atomic excitation, ${\bf k}$ is the spin-wave's wave-vector,  ${\bf x}_i$ is the position of the $i^{th}$ atom, $\ket{G}$ denotes the sate with all atoms in $\ket{g}$, and $N$ being the number of atoms in the ensemble.  The spatial profile of the collective excitations is considered in $C_{\{0/1\}}({\bf x}_i)$. Their shape is determined through the storage process and the shape of the input pulses.   %the control phase gate $CZ=2\ket{\overline{00}}\bra{\overline{00}}-\mathbb{I}$ operates
 
The {\it gate operation} is performed  by exciting the $\ket{\bar{1}}$ state of control and target to the Rydberg level via a global pulse.  
 The Rydberg excitation is applied by two-photon excitation with $\Omega_{1}$ and $\Omega_{2}$ Rabi frequencies that are red detuned from the  $\ket{5P_{1/2},F=m_f=1}$ intermediate state by $\Delta$ and tuned to $\ket{nD_{5/2},1/2}$ Rydberg state, see Fig.~\ref{Fig1}c. The $\sigma^+$ polarization of $\Omega_1$ laser would only excite the $\ket{1}=\ket{F=2,m_f=0}$ electronic state to the Rydberg level due to the selection rules, leaving the $\ket{0}$ state unchanged.  The effective gate Hamiltonian acting on the two spin-waves is
\begin{equation}
\label{Eq_H}
H=\sum_i [(\frac{\Omega_1(t)}{2}\sigma^{(i)}_{1p}+h.c.)+\frac{\Omega_2(t)}{2}\sigma^{(i)}_{rp}+h.c.) +\Delta \sigma^{(i)}_{pp}]
+\sum_{i<j}V({\bf x_i}-{\bf x_j})\sigma^{(i)}_{rr}\sigma^{(j)}_{rr}
\end{equation}
where $\sigma^{(i)}_{lm}=\ket{l}\bra{m}$ is the transition operator of the $i^{th}$ atom.

  \begin{figure} [h]
\centering 
       \scalebox{0.32}{\includegraphics{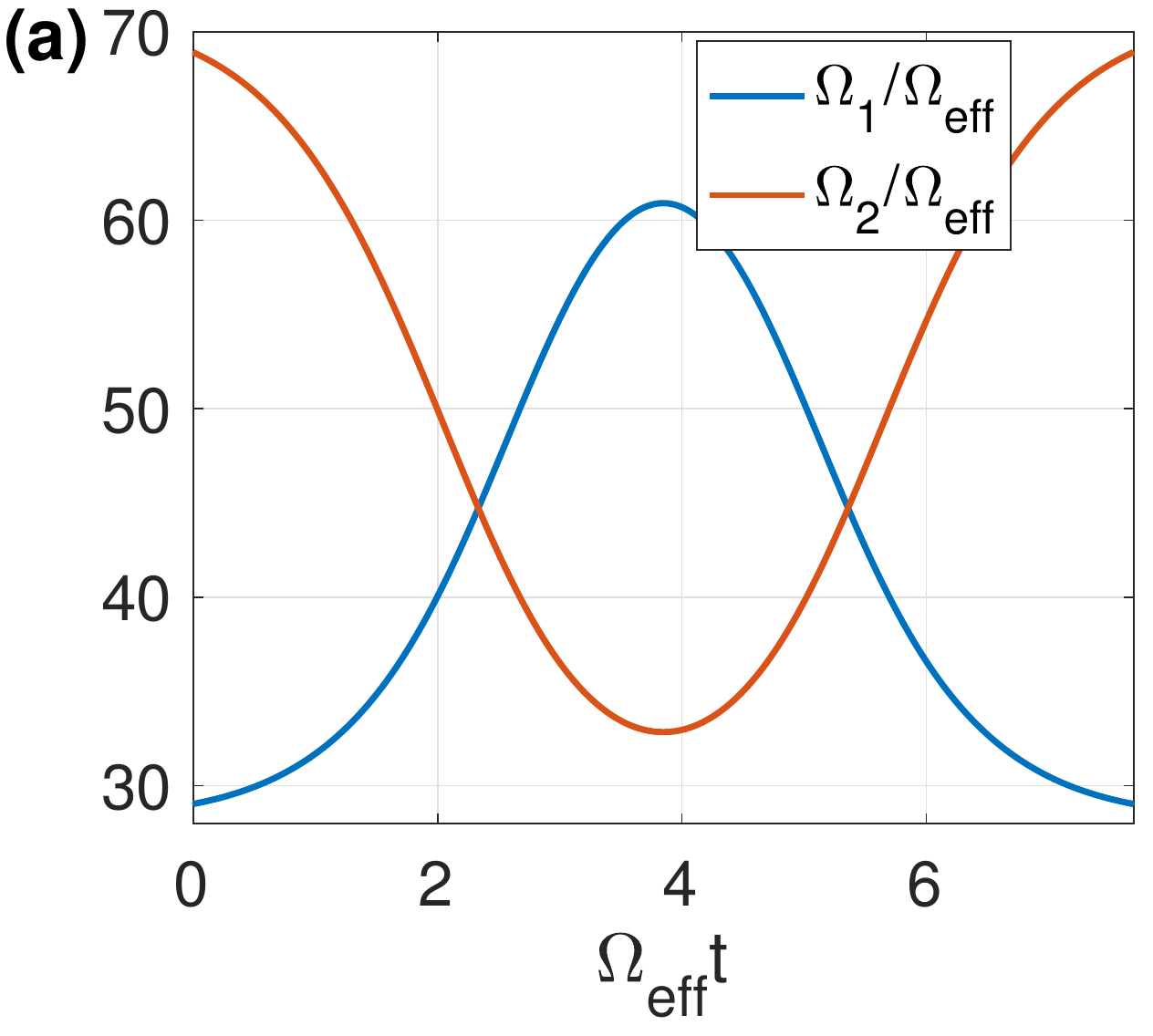}} 
       \scalebox{0.32}{\includegraphics{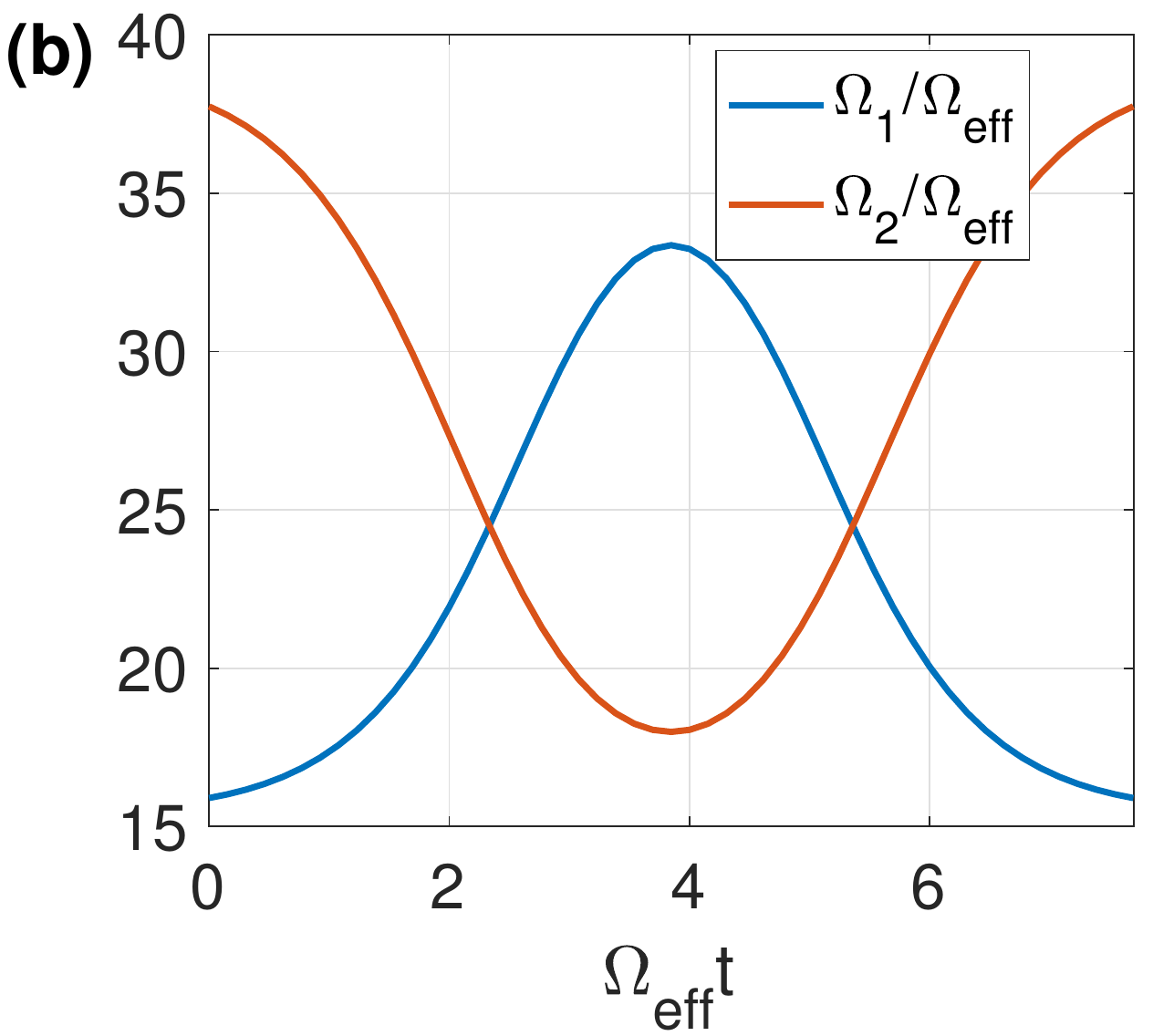}} \\
    %   \scalebox{0.32}{\includegraphics{Om1Om2}} 
      % \scalebox{0.32}{\includegraphics{Om1Om2_300}} 
   %    \scalebox{0.32}{\includegraphics{PrPhi01}} 
     %  \scalebox{0.32}{\includegraphics{PrPhi11}} 
     %  \scalebox{0.32}{\includegraphics{GaussianOm1Om2}} 
       \scalebox{0.32}{\includegraphics{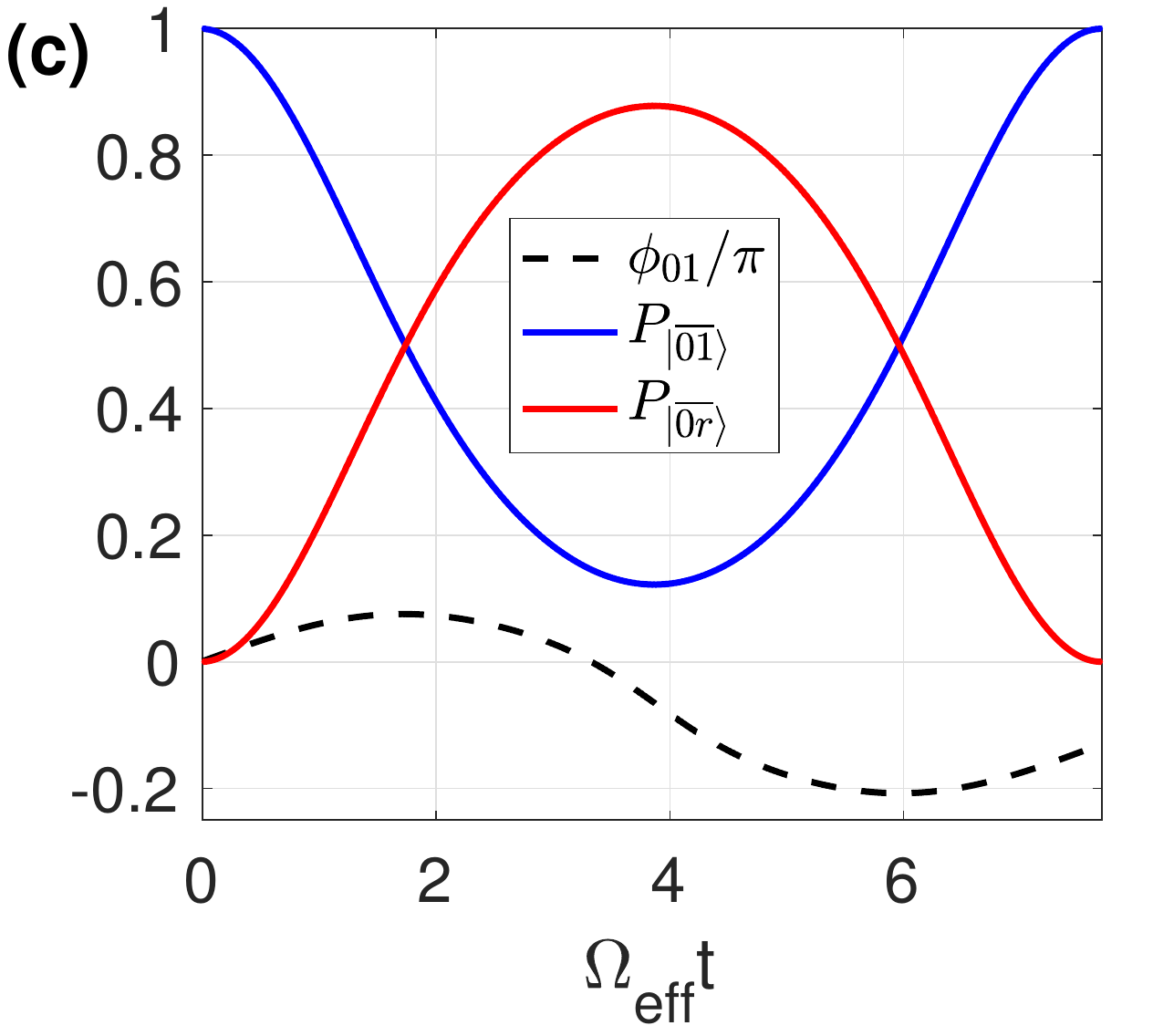}} 
       \scalebox{0.32}{\includegraphics{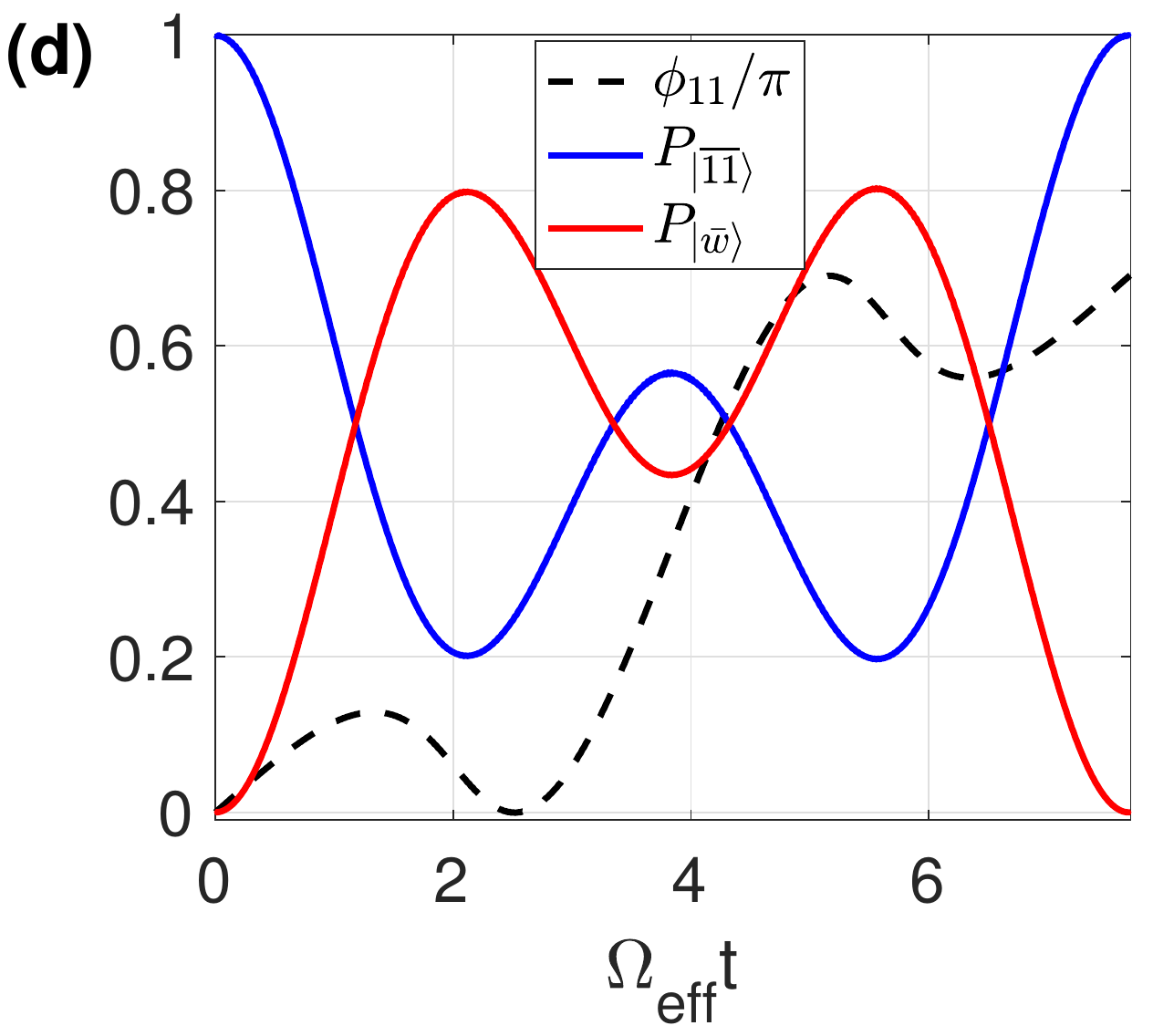}} 
 \caption{  Simulation of the gate operation (Eq.~\ref{Eq_H}) under a single continuous pulse.  Changing the Rabi frequencies $\Omega_{1,2}$ in the form plotted in (a,b) keeps the effective Rabi frequency constant  $\Omega_{\text{eff}}=\Delta/1000$ in (a) and  $\Omega_{\text{eff}}=\Delta/300$ in (b), and operates the CZ gate over $t_{\text{gate}}=7.7/\Omega_{\text{eff}}$. The optimized profiles in (a,b) are close to Gaussian profiles with $1/e^2$ width of $2.5/\Omega_{\text{eff}}$. (c) In cases of qubit configurations $|\overline{01}\rangle$ or $|\overline{10}\rangle$, the time evolution of the ground state population and the excited state $|\overline{0r}\rangle$ or $|\overline{r0}\rangle$ are plotted as well as the accumulated phase $\phi_{01}$. (d) For the qubit configuration of $|\overline{11}\rangle$, the time evolution of the ground $|\overline{11}\rangle$, $w=(|\overline{1r}\rangle+|\overline{r1}\rangle)/\sqrt{2}$ state, and the accumulated phase $\phi_{11}$ are plotted.   
  }\label{Fig2}
\end{figure}

 To apply the desired phase gate $CZ=2\ket{\overline{00}}\bra{\overline{00}}-\mathbb{I}$ with a single continuous global pulse, there are different dynamic parameters including the two lasers' amplitudes, frequencies, and phases that could be varied over time. Here only  the lasers' intensities get modulated for the sake of experimental simplicity. Figure \ref{Fig2} shows a suggestive Gaussian time evolution of the Rabi frequencies that retrieves the computational basis for all qubit configurations with the desired conditional phase arrangement.
 For the qubit configurations $\ket{\overline{10}}$ and $\ket{\overline{01}}$ the population rotation would add a phase of $\phi_{\overline{01}}=\phi_{\overline{10}}$, see Fig.~\ref{Fig2}c. 
 To describe the effects of relative laser intensities on the population rotation and acquired phase, a simplified case is discussed in App.~A.
 In the case of $\ket{\overline{11}}$, the presence of both spin waves within the blockade volume prevents double Rydberg excitation. The effective operation would then be in a two-dimensional space of $\ket{\overline{11}}$ and the excited $\bar{w}$ state $\ket{\bar{w}}=(\ket{\overline{1r}}+\ket{\overline{r1}})/\sqrt{2}$. As a result the system  derives with an enhanced Rabi frequency and a modified differential light-shift for the $\ket{\overline{11}}-\ket{\bar{w}}$ transition.  
Hence the population rotation and acquired phase would be different among distinguished qubit configurations as compared in Fig.~\ref{Fig2}c,d. To operate the $CZ$ gate, the relative Rabi frequencies $\Omega_1(t)/\Omega_2(t)$ would be varied over the operation to retrieve the qubit basis for all qubit configurations and to obtain the desired phase arrangement $\phi_{11}=2\phi_{01}-\pi$.
 Compensating for single qubit phase $\phi_{01}$ with a ground state light-shift, the desired $CZ$ operation will be obtained, see App.~C.
 Compared to conventional Rydberg blockade gate \cite{Jak00} with separate excitation sequence of the control and target qubits over $4\pi/\Omega_{\text{eff}}$ time, this gate operates at a shorter time interval of $t_{\text{gate}}=2.451\pi/\Omega_{\text{eff}}$ with $\Omega_{\text{eff}}=\Omega_1\Omega_2/2\Delta$. 
 %Remarkably, the Rydberg population over the gate operation is more than 2 times suppressed.
In contrast to the original proposal \cite{Jak00}, the Rydberg population over a 2$\pi$ pulse and over the entire gate operation is suppressed by 10\% and by two times respectively.
 This is an important factor as the short lifetime of Rydberg atoms is the main source of infidelity in Rydberg quantum computation. 

\section{ Fidelity \& Decoherence sources}

The main imperfection sources of the gate operation are caused by  spontaneous emissions from the Rydberg and intermediate levels, population rotation errors due to imperfect blockade, Doppler broadening of the transition line,  the loss of phase matching due to the thermal motion of atoms, and the inefficiencies of the photon storage in the atomic medium.  In Fig.~\ref{Fig3},\ref{FigX} the gate operation of Eq.~\ref{Eq_H} is solved numerically and the gate fidelity is averaged over all input qubit states. The operation is quantified by the phase-sensitive definition of fidelity \cite{Mol},
\begin{equation}
\label{FidDefenition}
F=[\text{Tr}(MM^{\dagger})+|\text{Tr}(M)|^2]/[n(n+1)]
\end{equation}
with $M=U_{id}^{\dagger}U_{gate}$, where $U_{id}$ and $U_{gate}$, represents ideal and realistic gate operations and $n=4$ is the dimension of qubit configurations.

\subsection*{Blockade efficiency}

\begin{figure} [h]
\centering 
       \scalebox{0.31}{\includegraphics{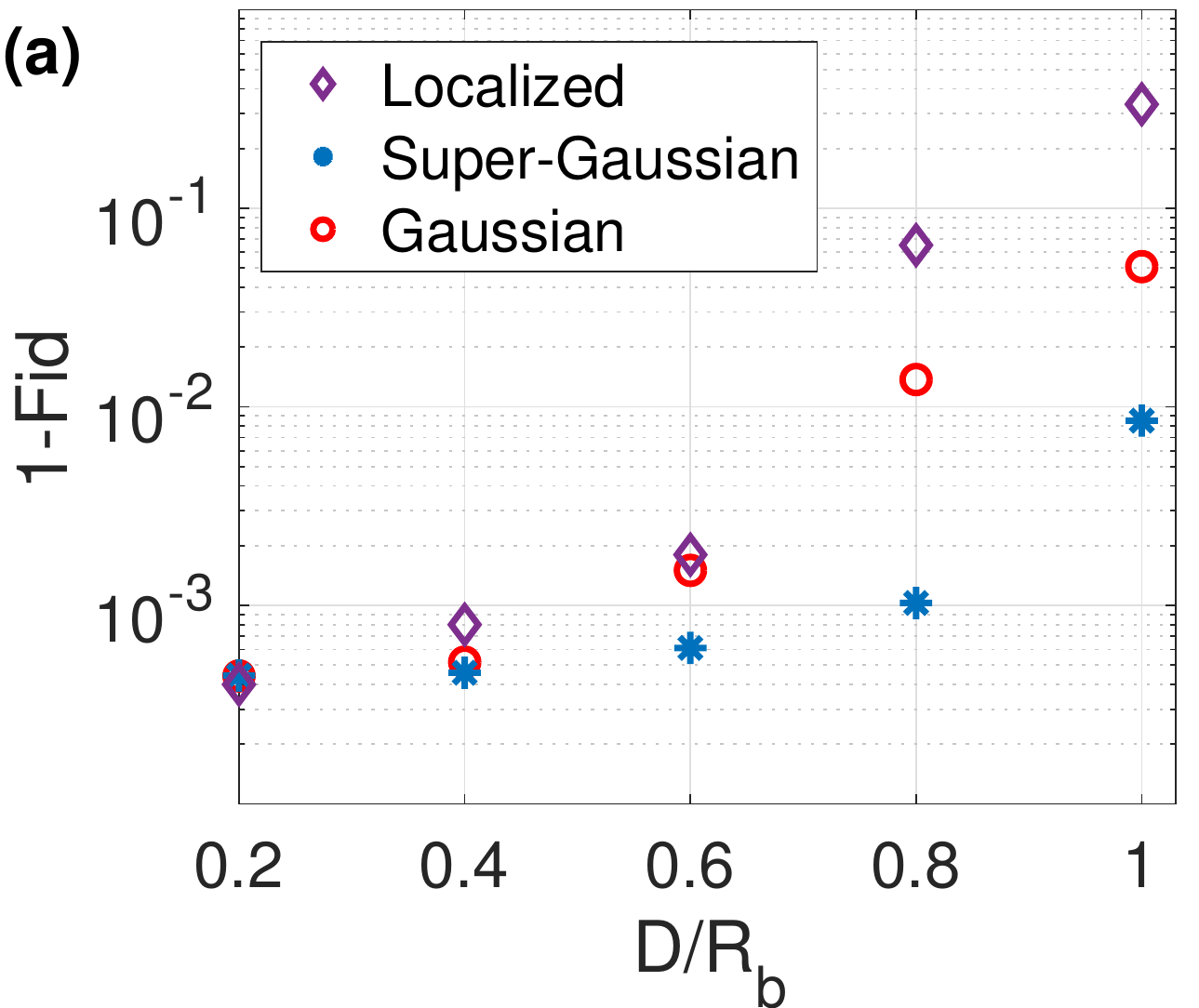}} 
       \scalebox{0.31}{\includegraphics{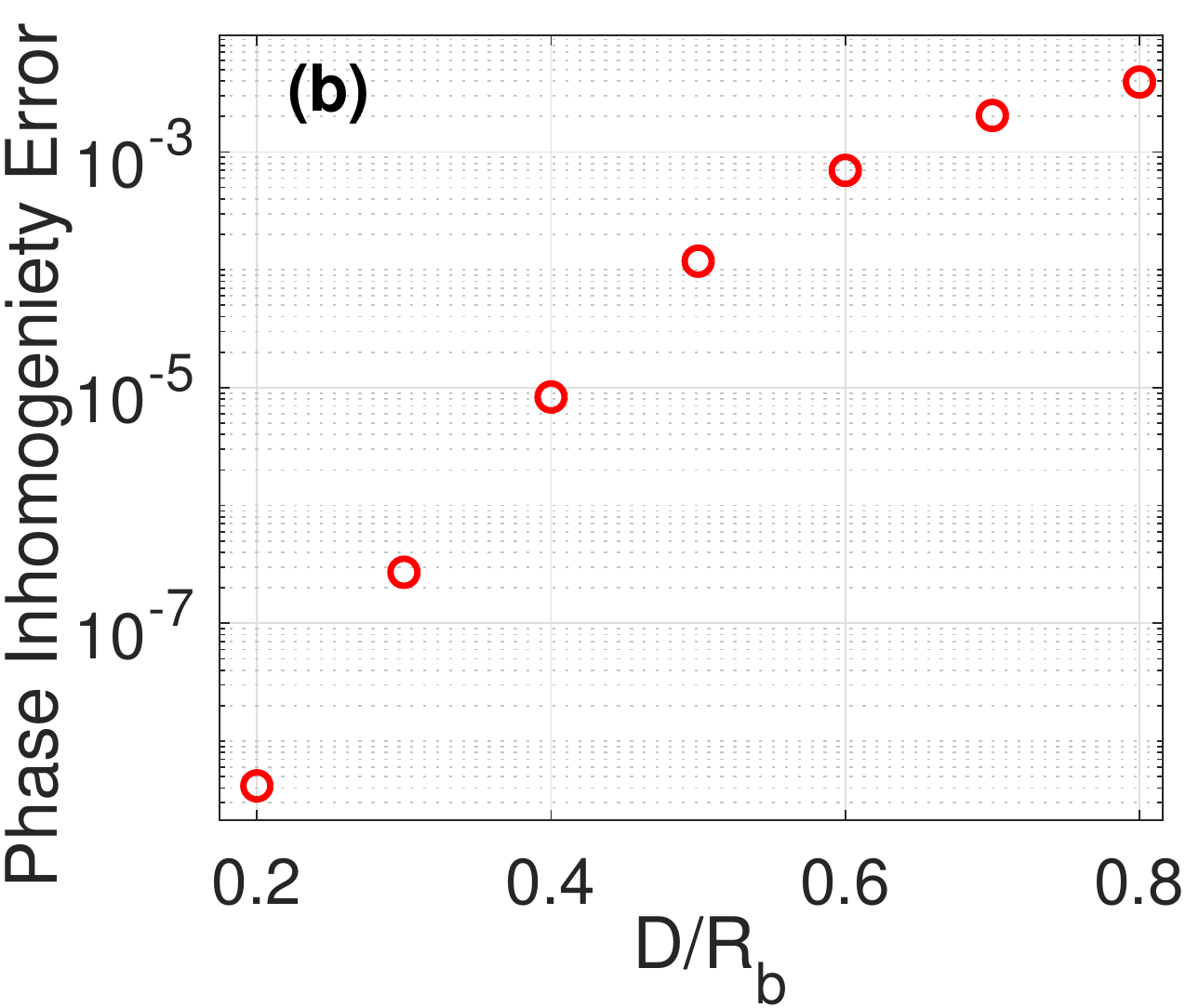}} 
              \scalebox{0.31}{\includegraphics{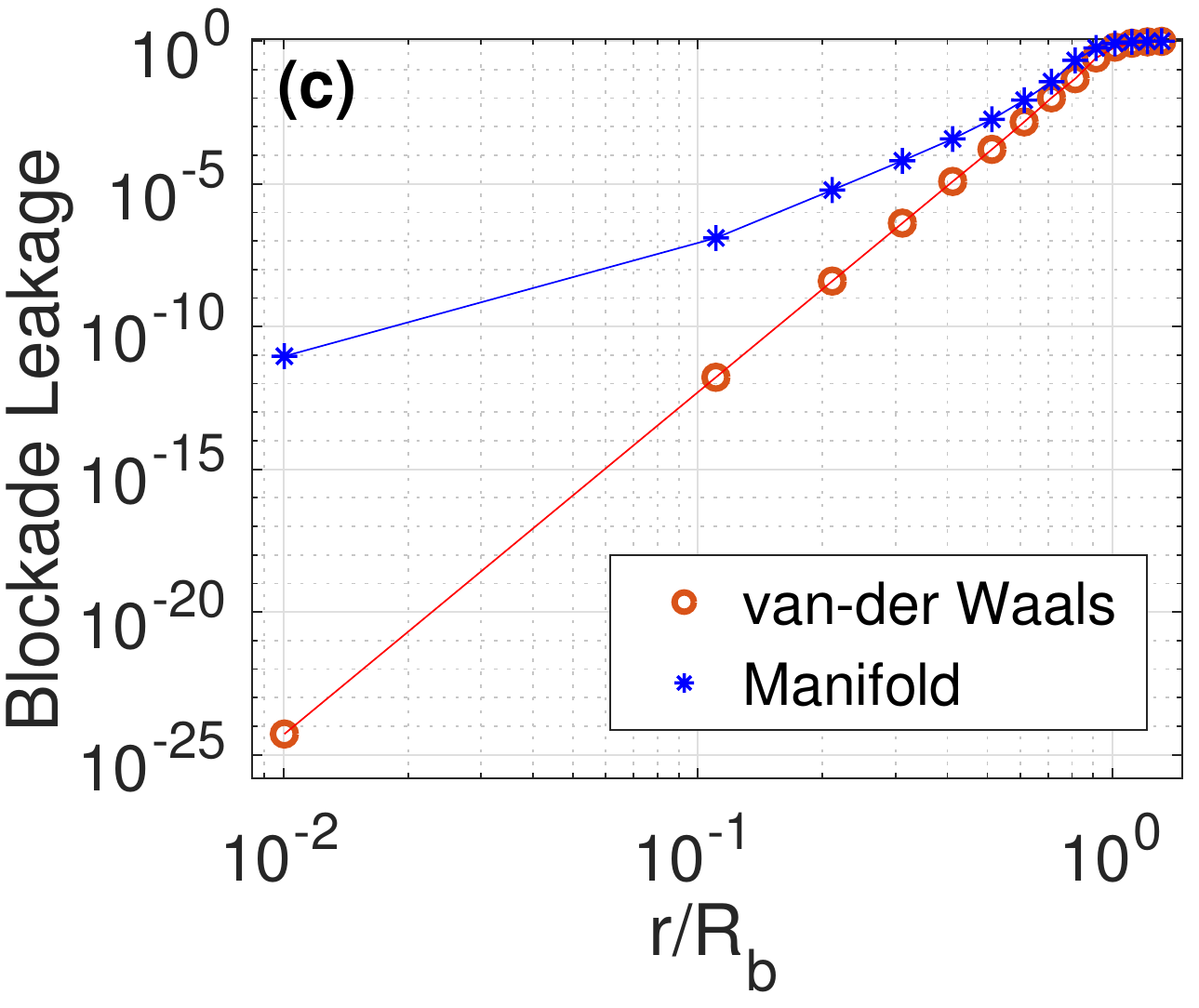}} 
\caption{Blockade imperfection. 
(a) Gate infidelity caused by an imperfect blockade between two localized atoms (diamond signs)  is plotted as a function of interatomic distance $D$ relative to blockade radius. 
In the case of delocalized excitations,  overlapping spin-waves with Gaussian (Circle sign) or Super-Gaussian (star signs) profiles are considered, see the text. The gate's infidelity is plotted as a function of the spin-waves' diameter  $D=2w$ at $1/e^2$ maximum. 
(b) The population leakage could impose an interaction-induced inhomogeneous phase over the spin-waves. Corresponding gate error is quantified as a function of spin-waves diameter normalized by blockade radius. 
(c) The validity of van-der Waals interaction applied in (a,b) is investigated. Here the maximum  blockade leakage over a 2$\pi$ rotation is quantified under a more realistic case where the targeted Rydberg pairs are dipole coupled to 1100 other neighbouring pairs (star signs). Compared to the simplified van-der Waals assumption (circle signs) the main deviation occurs at small interatomic distances where the strong couplings deviate the manifold's dynamics from the perturbative regime. 
For the desired case $0.6<D/R_b<0.8$ the error is determined by the double-excitation leakages that are separated far apart and hence the van-der Waals model gives a correct estimate of the gate's operation.   
}\label{FigX}
\end{figure}

Considering the space-dependent interaction over the two delocalized excitations, the {\it imperfect blockade} in large ensembles results in an inhomogeneous spin-rotation and rotation-induced phase acquisition over the spatial distribution of two spin-waves. 
In Fig.~\ref{FigX}a, the corresponding gate infidelity is quantified by Eq.~\ref{FidDefenition} for the case of localized qubits (diamond signs) as a function of their distance. The results are then extended to the case of two delocalized qubits in the form of spin-waves (circle/star signs) stored on top of each other.
 Here a 3-dimensional Gaussian    $|C_1({\bf x})|^2=\exp(-2x_{^{||}}^2/w_{^{||}}^2)\exp(-2{\bf x}_{\perp}^2/w_{\perp}^2)$ or super-Gaussian $|C_1({\bf x})|^2=\exp(-2x_{^{||}}^{10}/w_{_{||}}^{10})\exp(-2{\bf x}_{\perp}^{10}/w_{\perp}^{10})$ profiles are considered, with $x_{^{||}}$ and $x_{\perp}$ being the position vectors along and perpendicular  to the lasers' propagation direction. The gate's infidelity is plotted in Fig.~\ref{FigX}a as a function of the spin-waves' diameter  $D=2w$ at $1/e^2$ maximum. Considering the anisotropic van der Waals interaction, the same scaling with respect to blockade radius is considered in all directions $D/R_b=D_{||}/R_{b_{||}}=D_{\perp}/R_{b_{\perp}}$. Focusing the incoming photons with super-Gausian profile  to  $D=0.8R_b$ diameter ensures perfect operation with minor gate error of $10^{-3}$.

Blockade leakage could also cause an interaction-induced inhomogeneous phase. 
The phase acquired by each two interacting atoms $i$ and $j$ would be $\phi_{ij}=\int P_{rr}({\bf x_{ij}}) V({{\bf x_{ij}}}) \text{d}t$ where $V({{\bf x_{ij}}})$ is the distance dependent  van der Waals interaction with ${\bf x_{ij}}={\bf x_{i}}-{\bf x_{j}}$  and $P_{rr}({\bf x_{ij}})$ is the double-excitation leakage probability.
The inhomogeneous phase suppresses the  fidelity of $\ket{1_c1_t}$  state by
\begin{equation}
\label{Eq_LeakPhase}
E_{\text{IHPh}}=1-|\sum_{i,j}|C_1({\bf x_i})|^2 |C_1({\bf x_j})|^2 e^{-i\phi_{ij}}|^2.
\end{equation}
The corresponding gate error averaged over qubit configurations is plotted as a function of spin-waves' diameter $D=2w$ relative to the blockade radius in Fig.~\ref{FigX}b.
In terms of phase matching, since the leakage predominantly occurs at the edges of the spin-waves, this interaction-induced phase acts as a self-focusing term and hence does not increase photon loss.

While in the above discussion, a simplified van-der Waals interaction is considered, a more realistic case must encounter a huge manifold of dipole-coupled Rydberg pairs. This simplification is  justified here by investigating the blockade efficiency under the two models.
 Figure~\ref{FigX}c considers a manifold of 1100 strongly coupled Rydberg pairs with principal and angular quantum numbers being in the range of $n\in[98-102]$, $l\in[0-4]$. The maximum blockade leakage over a $2\pi$ pulse is plotted for the interatomic orientation parallel with laser fields. 
 The simulation is performed in two atom basis under the Hamiltonian 
% In two-atom basis, double excitation mechanisem subjected to the dipolar couplings in the Rydberg manifold is governed by
\begin{eqnarray}
H&=& \Omega/\sqrt{2}(|gg\rangle \langle r_0 g^{+}|+h.c.) + \Omega/\sqrt{2} (|r_0g^{+}\rangle \langle r_0 r_0|+h.c.)\\ \nonumber
&&+\sum_{i,j,k,l}\frac{C_{3_{ij,kl}}(\theta)}{R^3}(|r_l r_k\rangle \langle r_i r_j|+h.c.)+ \sum_{i,j}\delta_{ij}|r_i r_j\rangle \langle r_i r_j|
\end{eqnarray}
where the first line excites double Rydberg excitation via the intermediate symmetric state $\ket{r_0g^{+}}=(\ket{r_0g}+\ket{gr_0})/\sqrt{2}$. The dipolar interaction $C_{3_{ij,kl}}(\theta)/{R^3}$ mixes a manifold of Rydberg pairs $\ket{r_ir_j} $ that are detuned in energy from the initial targeted pair $\ket{r_0r_0}=\ket{100D_{5/2},1/2;100D_{5/2},1/2} $  by $\delta_{ij}$. While at large inter-atomic distances the result replicates the van-der Waals estimation $V\propto R^{-6}$, at  small interatomic distances the strong couplings deviate the manifold's dynamics from the perturbative regime. In quantifying the gate's operation on delocalized excitations in Fig.~\ref{FigX}a,b the dominant errors are caused by the excitations that are far apart and hence the van-der Waals model gives a correct estimate of gate's operation for the  interest range of spin-waves' size with $0.6<D/R_b<0.8$. 
Finally, at ultra-short distances in the presence of the strong level mixing, the blockade would be preserved due to the diluteness of the optically accessible Rydberg pairs in the  manifolds  \cite{Kha18}. This has also been proved by blockade experiments in dense atomic ensembles \cite{Dud12,Rip18,Ome20,Shi22}.

\subsection*{Other imperfection sources}

\begin{figure} [h]
\centering 
       \scalebox{0.36}{\includegraphics{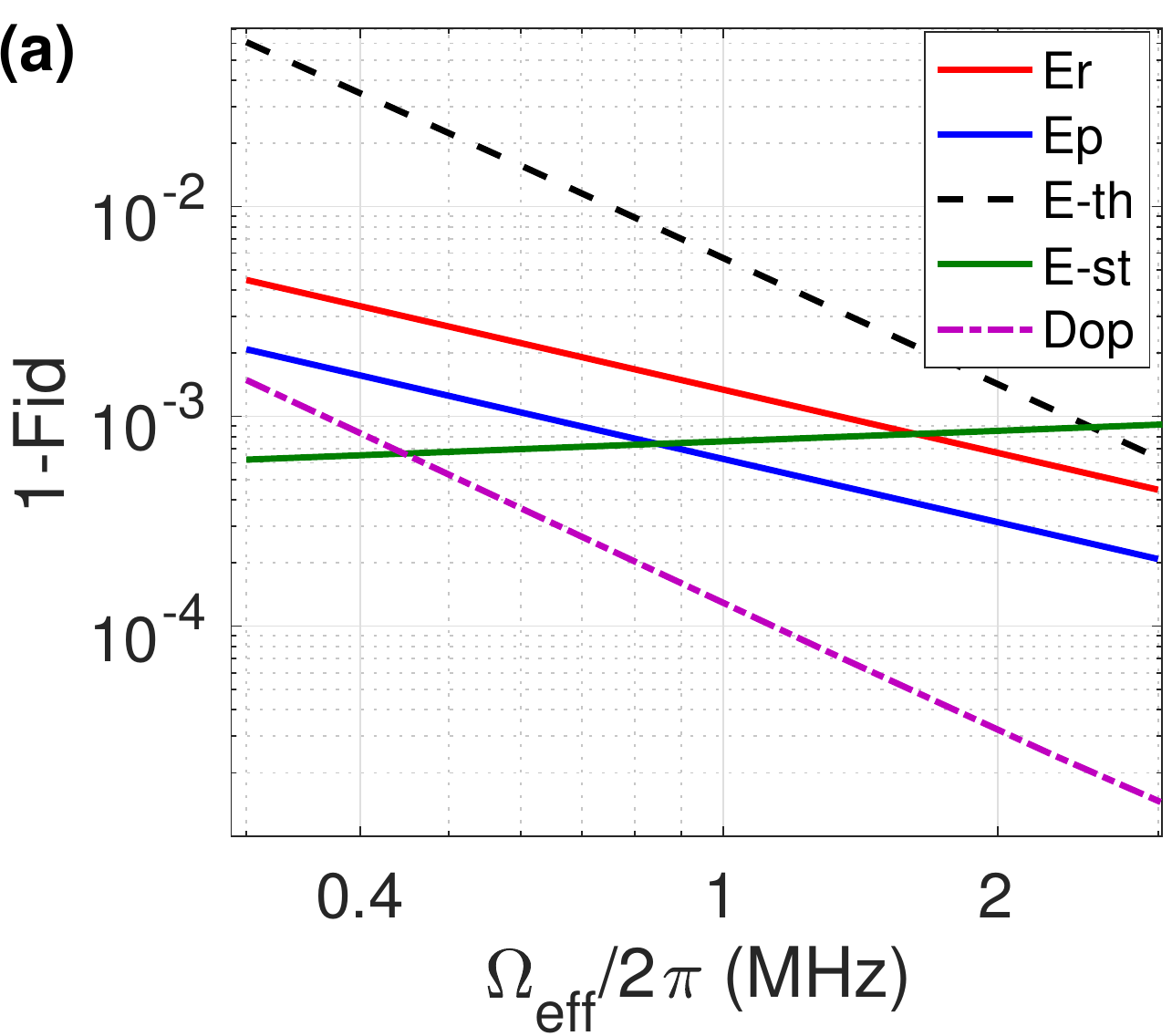}} 
       \scalebox{0.36}{\includegraphics{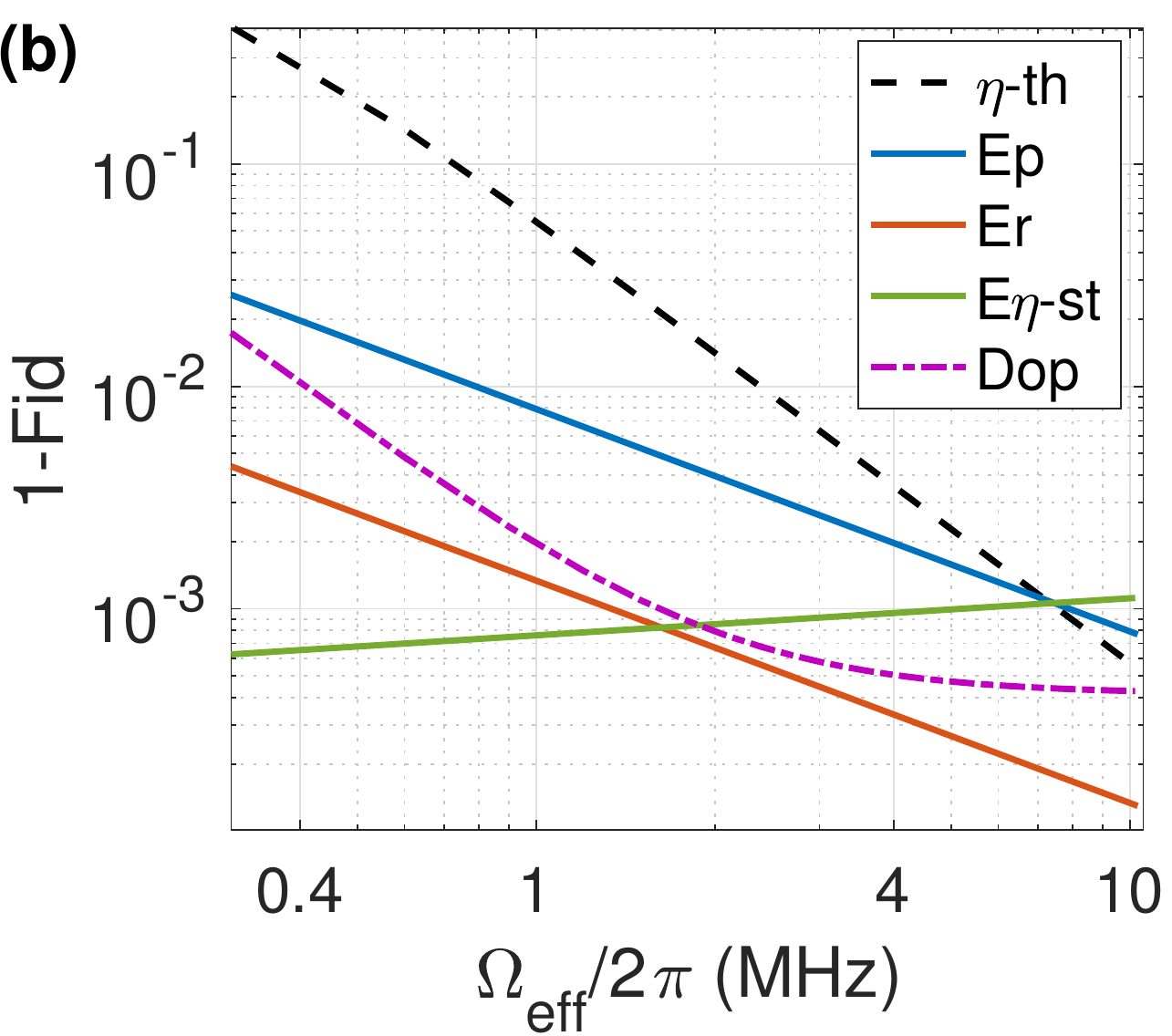}} 
\caption{Gate performance. 
 Error sources are quantified as a function of the gate speed in an ensemble cooled to (a) $T=1\mu$K and (b) $T=10\mu$K. Quantified error sources include spontaneous emission from the intermediate $p$ state $(E_p)$ and Rydberg levels $(E_r)$. Also, the  atomic thermal motion causes the loss of spin-waves' coherence $(E_{th})$ and raises the population rotation errors due to the Doppler frequency shift (Dop). The storage and retrieval inefficiency $E_{st}$ (under non-Rydberg EIT) is also quantified in an ensemble of length $0.8R_b$ and density of $10^{14}$cm$^{-3}$.  Here faster gate operation reduces the blockade radius and hence the medium length, which adversely affecting storage efficiency.  
Ultra-cold temperature allows slow operation and hence larger detuning from the intermediate state (a) $\Delta/\Omega_{\text{eff}}=1000$ and (b) 300 which suppresses the loss from the intermediate state. The target Rydberg state is $|100D_{5/2},1/2\rangle$ excited via $|6P_{1/2}\rangle$ intermediate states.
  }\label{Fig3}
\end{figure}

The   Doppler shift experienced by atoms moving with thermal velocity $v$ is $\delta_D={\bf v}.{\bf k}$ with ${\bf k}={\bf k}_1-{\bf k}_2$ being the effective wave-vector of the laser excitation. Hence, a {\it Doppler broadening} would be experienced by an ensemble with Maxwell velocity distribution $P_v(v)=\sqrt{m/2\pi KT}\exp(-mv^2/2KT)$, where $m$, $K$, and $T$ are the atomic mass, Boltzmann constant, and the ensemble temperature respectively. 
In an ensemble cooled to  1$\mu$K (10$\mu$K),  counter-propagating beam configuration with wavelengths 1006 nm and 421.7 nm results in a  Gaussian Broadening width of 2$\pi\times$19 kHz (2$\pi\times$61 kHz).
At low Rabi frequencies, this effective random detuning $\delta_{\text{D}}\sigma_{rr}$ applies significant deviation of phase and population that affects the average gate's fidelity quantified by Eq.~\ref{FidDefenition}. At strong driving regimes, the effect of Doppler line broadening would be negligible, see Fig.~\ref{Fig3}a,b. The lower limit of the Doppler broadening in Fig.~\ref{Fig3}b  is due to the large intermediate detuning $\Delta/\Omega_{\text{eff}}=300$ which is a limiting factor in population rotation. In case of a larger detuning $\Delta/\Omega_{\text{eff}}=1000$ of Fig.~\ref{Fig3}a the Doppler error keeps reducing at stronger driving regimes.

The {\it spontaneous emission} from the Rydberg level $\ket{100D_{5/2},1/2}$ ($\tau_r=343\mu$s) \cite{Bet09} and the intermediate state $\ket{6P_{1/2}}$ ($\tau_p=129$ns) suppresses the gate's fidelity as plotted in Fig.~\ref{Fig3}. 
 Here the loss is quantified phenomenologically by numerical calculation of the Schr\"odinger equation associated with the Hamiltonian of Eq.~\ref{Eq_H}. 
To obtain a pessimistic limit of infidelity the entire scattering is considered as loss. Slower operation in (a) allows larger detuning from the $\ket{p}$ state and hence suppresses the corresponding scattering rate. This error could be further suppressed by using the long-lived intermediate state of $\ket{4D}$. The Rydberg decay could be controlled by suppressing the blackbody radiation induced loss using a cryogenic environment or by exciting higher principal quantum numbers $\tau_{r}\propto n^3$.  

The {\it thermal motion} of atoms disturbs the phase-matching condition and washes out the spin-waves' coherence. 
The thermal motion's destructive effects quantify as $\eta_{_{th}}=\frac{1}{(1+(\frac{t}{\xi})^{2})^{2}}\exp[\frac{-t^{2}/\tau^{2}}{(1+(t/\xi)^{2})}]$
\cite{key-1} where $\tau=\frac{\Lambda}{2\pi v}$ is the de-phasing time scale corresponding to the atomic motion over a spin-wave's wave-length
$\Lambda$ with the thermal speed $v$, and $\xi=\frac{w}{v}$ is the time scale that an atom moves out of spin-wave's profile width $w$ that is about the same length as blockade radius.
For the temperature of $T=1\mu$k and $10\mu$k the thermal speed of atoms along one direction would be $v_{th}=\sqrt{K_bT/m}=0.01$ and $0.03$m/s with $K_b$, $m$ being the Boltzmann constant and mass of Rb atoms. 
Considering the non-Rydberg EIT storage scheme of Fig.~\ref{Fig1}a with counter-propagating photon and classical field, the spin-wave's wavelength $\Lambda=10$cm is much larger than the blockade radius and could be neglected. However considering the Rydberg excited spin-wave $\ket{\bar{r}}=\sum e^{i{\bf k_r.x_i}}\ket{g^1...r^i..g^N}$ with $k_r=k_{1006}-k_{421}$, the excitation's wave-length would be $\Lambda=714$nm which is much shorter than the blockade radius and acts as the limiting factor. 
Fig.~\ref{Fig3}a and \ref{Fig3}b quantifies the scale of thermal motion infidelity as a function of the gate speed  in  an ultra-cold medium with  $T=1$ and $10\mu$K. The thermal speed of atoms could be suppressed by further cooling or  deploying heavier atoms like Cs. The  Rydberg wave-vector could be enhanced to $\Lambda=4.5\mu$m by using $\ket{5D}$ intermediate state in Cs ensemble.

The {\it storage and retrieval efficiency} in the free space EIT is given by $\eta=1-1/OD$ \cite{Gor07} where the optical depth is given by  $OD=N\Lambda^2/A$ with $N$ being the number of atoms, $\Lambda$ the carrier wavelength of the photon, and $A$ the cross-section. In a spherical ensemble with radius $R_b$ we have  $OD=4/3\rho R_b\Lambda^2$. In Fig.~\ref{Fig3}a,b the storage efficiency is plotted as a function of the gate speed. Higher Rabi frequencies reduce the blockade radius $R_b=(C_6/\Omega_{\text{eff}})^{1/6}$ and hence the optical depth, see Fig.~\ref{Fig3}a,b. Exciting higher Rydberg principal numbers $n$ could enhance the blockade radius $R_b\appropto n^{2}$ and the optical depth. Cavities could also enhance the optical depth by a factor of $F/\pi$ where $F$ is the cavity finesse \cite{Jia19}.

In an idealized case, we assumed constant spatial and temporal Rabi frequency. In a realistic case, the effects of {\it  inhomogeneity and stability} of exciting beams must be taken into account \cite{Day22}. 
In an optical transition $\Omega\propto\sqrt{I}$ with $I $ being the laser intensity,  the Rabi frequency fluctuations due to intensity noise is given by $\delta\Omega/\Omega=\delta I/2I$. In a two-photon excitation $\Omega_{\text{eff}}\propto\sqrt{I_1}\sqrt{I_2}$ the Rabi frequency encountering the fluctuations would be given by 
 $\Omega_{\text{eff}}(1+\delta I_1/2I_1+\delta I_2/2I_2)$. The noise induced infidelity over a $2\pi$ pulse would be $1-Fid\approx \pi^2(\delta I_1/2I_1+\delta I_2/2I_2)^2$. Considering the RMS intensity noise of  0.1\% the corresponding infidelity would be $10^{-5}$.
Regarding the spatial homogeneity, a super-Gaussian transversal profile could be considered for the single-photons (sp) and the classical lasers (L) with HW at $1/e^2$ maximums of $w_L=1.3R_{b \perp}$ and $w_{sp}=0.8R_{b \perp}$. In this arrangement, the gate infidelity of $2\times 10^{-4}$ caused by the laser spatial inhomogeneity must be encountered. 
Using Gaussian profiles  for the single-photons ($w_{sp}=0.8R_{b \perp}$) and driving lasers ($w_L=2R_{b \perp}$) the  infidelity of $6\times 10^{-4}$  would contribute to the gate error budgeting.

 The Optimum gate speeds discussed in Fig.~\ref{Fig3}a [\ref{Fig3}b]  are $\Omega_{\text{eff}}/2\pi=2.3$MHz [7MHz]. Considering the  super-Gaussian cross-section $I=I_0e^{-2({\bf \rho}/w_{\perp})^{10}}$  \cite{Gil16},   with the half-width of $w_{\perp}=1.3R_{b \perp}=7.6\mu$m [ 6.3$\mu$m], the required laser powers for the 1006nm laser is 230mW [310mW], see App.~B. The transition matrix elements in $^{87}Rb$ are given by $\langle r\rangle_{5S}^{6P_{3/2}}=0.528a_0$, $\langle r\rangle_{5S}^{6P_{1/2}}=0.235a_0$ , and $\langle r\rangle_{6P}^{nD}=0.035\times(53/n)^{3/2}a_0$.  The large dipole transition of 5S-6P allows the same order of Rabi frequency  for the 421.7nm transition with a 1mW laser. 
The mentioned gate speeds provide high-fidelity operations of 99.7\% with available experimental parameters.

{\bf Conclusion:}
The proposed photonic gate scheme stores photons in an atomic ensemble and excite them to Rydberg levels in separate steps. The gate operation is carried out via a single continuous global pulse. The desired operation is performed by dynamic modulation of laser intensities applied in the two-photon excitation in the presence of dipolar interaction. This scheme closes major decoherence channels of previous photonic gates and therefore, results to high fidelity operations.

{\bf Competing interests-}
The author declares no conflicts of interest.

{\bf Data availability.} Data underlying the results presented in this paper are not publicly available at this time but may be obtained from the authors upon reasonable request.

\section*{Appendix A}

To understand the effects of relative laser intensities on the population rotation and acquired phase, here a simplified case with constant laser parameter is discussed analytically.
In the regime of large detuning, $\Delta\gg \Omega_{1,2}$ the intermediate state would be eliminated adiabatically leading to an effective two-level system. Under constant laser parameters, the stored state  $\ket{\overline{01}}$  would evolve as
 \begin{figure} [h!]
\centering 
       \scalebox{0.38}{\includegraphics{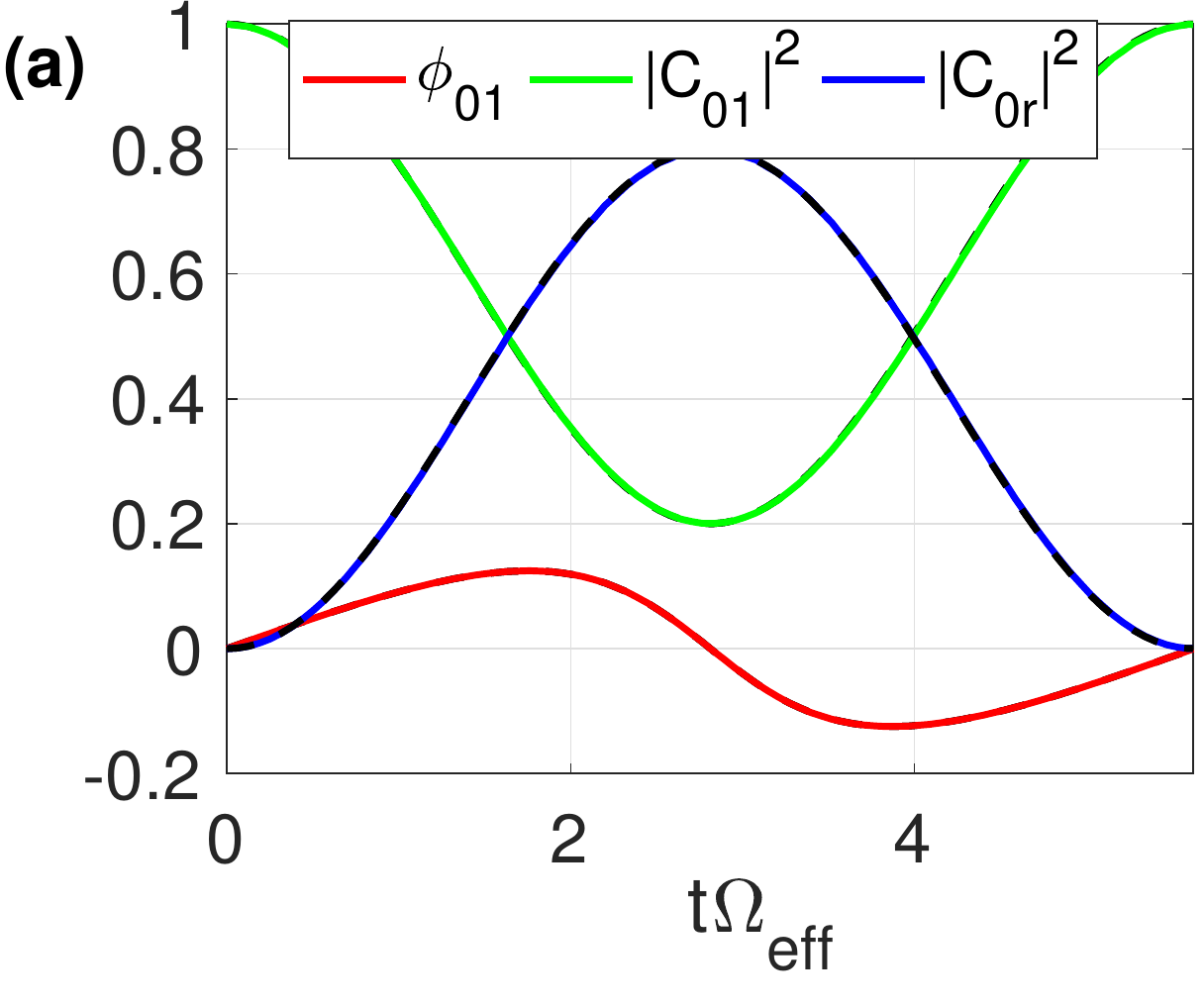}} 
       \scalebox{0.38}{\includegraphics{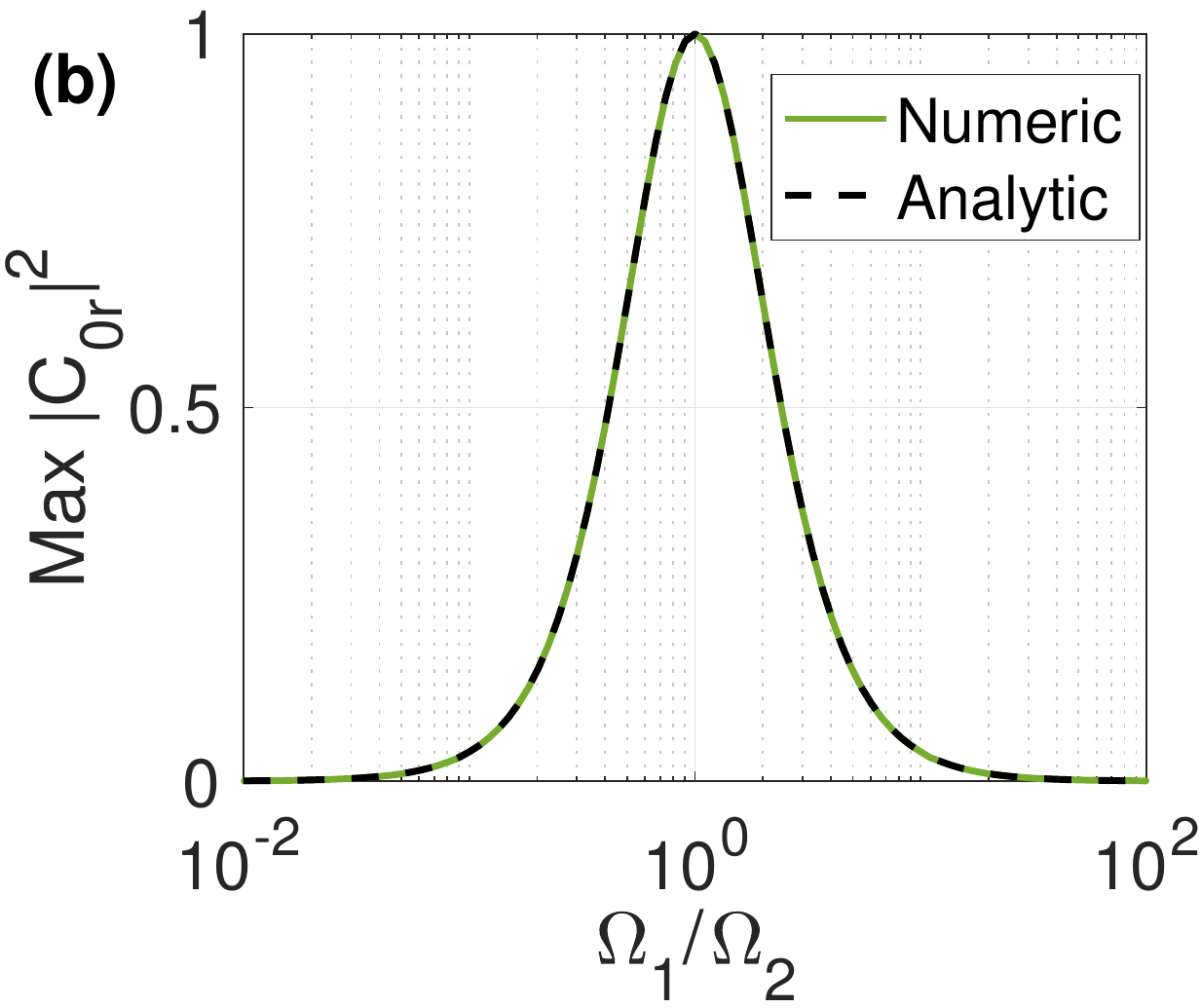}} 
\caption{Justifying the analytic model of Eq.~\ref{Eq_Analytic} (dashed lines) with the numerical simulation of Eq.~\ref{Eq_H} (solid lines). (a) Time evolution of the Rydberg excitation of $|\overline{01}\rangle$ qubit state with laser parameters  $\Omega_1/\Omega_2=0.5$, $\Delta/\Omega_{\text{eff}}=1000$, $\Omega_{\text{eff}}/2\pi=10$MHz.  (b) The maximum population of the Rydberg state $|C_{0r}|^2$ could be controlled by adjusting the relative intensity of lasers $\Omega_1/\Omega_2$. }\label{Fig4}
\end{figure} 
\begin{eqnarray}
&&\ket{\psi(t)}= C_{01}(t)\ket{\overline{01}}+C_{0r}(t)\ket{\overline{0r}}\\ \nonumber
&&C_{01}(t)=\frac{\Omega_2^2+\Omega_1^2 \text{e}^{\text{i}t\frac{\Omega_1^2+\Omega_2^2}{4\Delta}}}{\Omega_1^2+\Omega_2^2}\\ \nonumber
&&C_{0r}(t)=\frac{2\Omega_1\Omega_2}{\Omega_1^2+\Omega_2^2}\text{e}^{\text{i}t\frac{\Omega_1^2-\Omega_2^2}{8\Delta}}\sin(\frac{\Omega_1^2+\Omega_2^2}{8\Delta}t)\\ \nonumber
\label{Eq_Analytic}
\end{eqnarray}
The population of $\ket{\overline{0r}}$ state and the acquired phase depends on the ratio of $\Omega_1/\Omega_2$ as can be seen in Fig.~\ref{Fig4}.

\section*{Appendix B: Achievable Rabi frequencies}
Here we discuss the achievable Rabi frequencies for the desired super-Gaussian beams. For the 6P-nD transition strong laser powers above 100W are available in pulsed lasers \cite{Vri20}. For a super-Gaussian cross-section profile, the electric field at the center of the beam $E_0$ could be obtained from  $P=E_0^2/2\eta \int  \exp(-2 r^{10}/w^{10}) 2\pi r \text{d}r$, where $\eta$=377, w=7.6$\mu$m, P=100W. 
The transition matrix elements in $^{87}Rb$ are given by $\langle r\rangle_{5S}^{6P_{3/2}}=0.528a_0$, $\langle r\rangle_{5S}^{6P_{1/2}}=0.235a_0$ , and $\langle r\rangle_{6P}^{nD}=0.035\times(53/n)^{3/2}a_0$.
Hence, the maximum achievable Rabi frequency  would be  $\Omega_2=eE_0\times \langle r\rangle_{6P}^{nD} /\hbar=2\pi \times 4$GHz. In the paper,  a 20-fold smaller Rabi frequency is considered for optimum operation, resulting in a less demanding experiment.

\section*{Appendix C: single-atom phase correction}
Over the gate operation with 421.7nm and 1006nm lasers, the former laser applies a differential light shift to the qubit levels. Also, the gate generates $\phi_{01}$ on each atom at $|1\rangle$ qubit states. These single-atom phases could be compensated after the CZ gate operation by applying a qubit $X(\pi)$ rotation on all atoms and then applying the 421nm pulse (without 1006nm) to correct single qubit phases. The qubits would then be rotated back by a $-X(\pi)$.

\end{document}